\documentstyle[epsf,prl,aps]{revtex}
\begin{document}
\twocolumn[\hsize\textwidth\columnwidth\hsize\csname 
@twocolumnfalse\endcsname 
\title{Role of defects in the electronic properties
of amorphous/crystalline Si interface}  
  \author{Maria Peressi,$^{a,b,\ast}$\footnotemark[0] Luciano Colombo,$^{a,c}$
          and Stefano de Gironcoli$^{a,d}$}
  \address{ $(a)$ Istituto Nazionale per la Fisica della Materia \\
            $(b)$ Dipartimento di  Fisica Teorica, Universit\`a di
                  Trieste, Strada Costiera 11, I-34014 Trieste, Italy\\
            $(c)$ Dipartimento di  Fisica, Universit\`a di Cagliari,
                  Cittadella Universitaria, I-09042 Monserrato (CA),
                  Italy\\
            $(d)$ Scuola Internazionale Superiore di Studi Avanzati, 
                  via Beirut 2-4,  I-34014 Trieste, Italy
          }
\date{\today} 
\maketitle

\begin{abstract}
The mechanism determining the
band alignment of the amorphous/crystalline Si heterostructures is 
addressed with direct atomistic simulations
of the interface performed using a hierarchical combination of various 
computational schemes 
ranging from classical model-potential molecular 
dynamics to ab-initio methods.
We found that in coordination defect-free samples
the band alignment is almost vanishing and independent on interface 
details.
In defect-rich samples, instead, the band alignment is sizeably different 
with respect to the defect-free case, 
but, remarkably, almost independent on the
concentration of  defects.
We rationalize these findings within the theory 
of semiconductor interfaces.
\end{abstract}
]
\footnotetext[0]{$^\ast$ E-mail: \tt peressi@ts.infn.it}
Amorphous-crystalline silicon interfaces (a-Si/c-Si and a-Si:H/c-Si)
are of unquestionable technological importance for their use  in solar 
cells 
and in other optoelectronic devices. Their electronic and optical 
properties 
are governed by the band discontinuities \cite{Fantoni,Song,Gall}, whose 
value and origin however have not
been clearly established so far. To our knowledge, few and conflicting
experimental results exist for  a-Si:H/c-Si, 
ranging from 0 to 0.7 eV for the valence band offset (VBO) 
(see Refs.\cite{Fantoni,Song,Gall,Evangelisti,VdW} 
and references therein), and even less 
informations are available for  a-Si/c-Si \cite{Ley}.
On the theoretical side, the problem of band alignment
is addressed in two papers  \cite{VdW,Allan}, but only from a comparison of
the two phases separately and not from a direct simulation of the 
interface.
The atomic-scale modelization of the interface and its structural 
properties 
are the subject of few other works \cite{Saito,Kwon,Kaxiras,Tersoff} which, 
however, do not examine the electronic structure.
The combined study of the structural and electronic properties is necessary
to understand the mechanism determining the band alignment and in 
particular
the role of bulk-related and  interface-specific details.

With this aim we address here our attention 
to the a-Si/c-Si junction, described using periodically repeated 
supercells.
To generate the structures
we adopted a hierarchical combination of models and simulation schemes,
including the well known Wooten-Winer-Weaire (WWW) model \cite{WWW},
model--potential  \cite{EDIP}
and semi--empirical tight--binding (TB) molecular dynamics 
(MD) \cite{TB-MD}. We used then
 ab initio state-of-the-art pseudopotential calculations 
\cite{PWSCF,noi-aSi} to refine the equilibrium geometries and 
to compute the electronic structure.

We will consider a-Si/c-Si structures with
different a-Si atomic-scale morphologies.
In order to investigate the bulk-related effects, 
we will also study for comparison 
the band lineup at crystalline strained/unstrained Si homojunctions
 and at heterocrystalline cubic diamond/hexagonal diamond
interfaces.

{\it Atomic-scale model for a-Si/c-Si interface.} --
For the description of the (001)-oriented a-Si/c-Si interface we
consider periodically repeated tetragonal
(001) supercells containing 320 Si atoms, with cell dimensions
$a=2a_0$ and $c\approx 10a_0$, where $a_0$ is the lattice parameter of
the crystalline phase.
The analysis of the electronic structure, in
particular of the spatial decay in the crystalline region
of the  gap states induced by coordination defects of the a-Si region,
shows that this  cell size is large enough to recover crystalline bulk
features in the middle of the c-Si slab. 

The atomistic models of the interface
structure are obtained according to a procedure that, basically,
consists in the following four steps:
$(i)$ a tetragonal cell of a-Si with dimensions
$a=2a_0$, $c=5a_0$ and 160-atoms
is created by quenching from the 
melt during a constant--pressure MD simulation based on the 
``environment dependent interatomic potential'' (EDIP) ~\cite{EDIP}; 
the quenching procedure was operated by velocity rescaling at a rate
as slow as 0.1 K/ps (i.e. about 100 times slower than typical TB--MD
simulations~\cite{servalli}), so that density fluctuations during the
liquid--to--amorphous transition were properly damped and a fully
stress--free a-Si sample was indeed obtained;
a sample interface is then created by joining 
through the smallest face the above a-Si sample to a c-Si one
with the same cell;
$(ii)$ the structure is further annealed up to 800 K
(heating rate $\sim$ 1 K/ps), aged at that temperature for about
100 ps, and finally relaxed down to 0 K (cooling rate $\sim$ 1 K/ps)
by means of constant--pressure, constant--temperature EDIP--MD simulations;
$(iii)$
at this point we switch from EDIP--MD to TB--MD, adopting the
parametrization described in Ref.~\cite{TBmodel}; since the two models
give different equilibrium lattice parameter for c-Si ($a_0^{EDIP}$=5.43 
\AA\,
to be compared with $a_0^{TB}$=5.45 \AA\ ), we first
perform an homogeneous rescaling of the 
supercell dimensions and atomic positions, thus  keeping the
optimal $c/a$ ratio provided by the previous two steps; $(iv)$
the {\em internal} degrees of freedom are then further relaxed by means
of a new constant--volume TB--MD annealing procedure (annealing temperature
is 1000 K, the heating/cooling rates are, respectively, 500 K/ps and 
160 K/ps).
Overall the computational workload is heavy - especially for the TB--MD steps 
- and has to be mastered by means of O(N) formulation of TB--MD 
and/or  extensive use of distributed computing ~\cite{TB-MD}.
 
\begin{figure}[h]
\epsfclipon
\epsfxsize=8cm
\centerline{
\epsffile{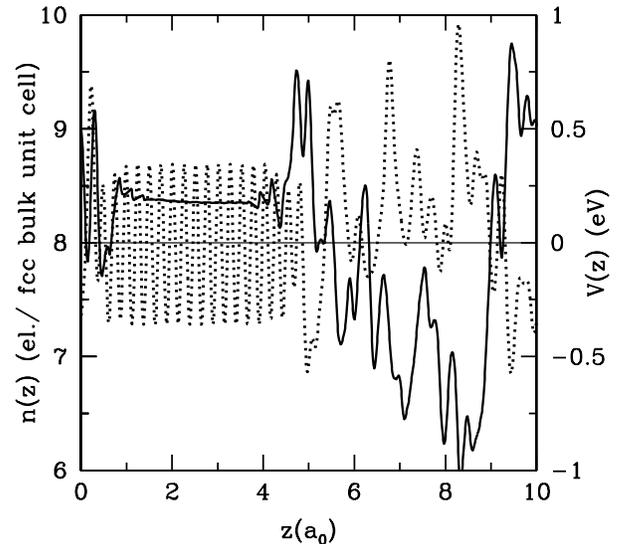}
}
\caption{Profile of the electron density (dotted line,
left scale) and of the macroscopic average of the total electrostatic
potential (solid line, right scale) for the sample $D$ (see text). 
The position along the growth
direction $z$ is indicated in units of $a_0$, the lattice parameter
of the crystalline phase.}
\label{f:potential}
\end{figure}

\begin{figure}[h]
\epsfclipon
\epsfxsize=8cm
\centerline{
\epsffile{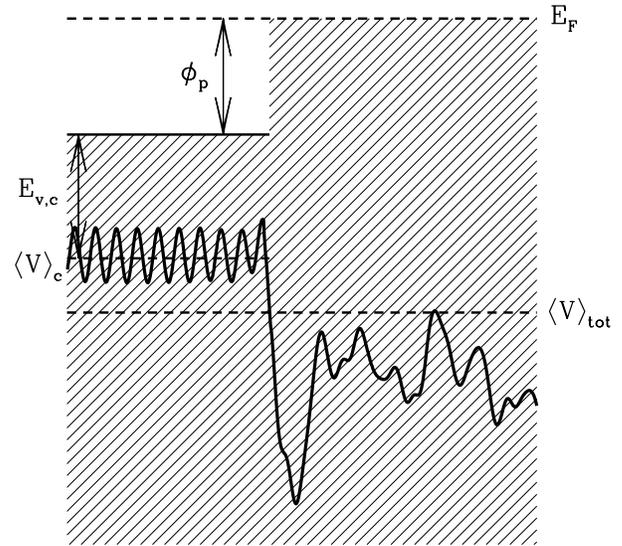}
}
\caption{Scheme of the
band alignment at a-Si/c-Si interface with defect-rich a-Si, with
indication of:
a typical profile of the electrostatic potential in the supercell
(thick solid line), its average over the whole supercell
($\langle V\rangle_{tot}$), its average in the crystalline region
($\langle V\rangle_c$), 
the Fermi energy of the supercell ($E_F$), the 
valence band top edge of c-Si ($E_{v,c}$) with
respect to $\langle V\rangle_{tot}$, and finally
the Schottky barrier $\phi_p$.} 
\label{f:sb}
\end{figure}

The resulting structure is then used
as input of ab initio  calculations, performed within 
the framework of  density-functional theory in the local density
approximation, using norm-conserving pseudopotential \cite{Gonze},
a plane-wave cutoff of 16 Ry, and 4 {\bf k} points generated by
a symmetrized Monkhorst-Pack grid \cite{MP}.
We first perform a further homogeneous rescaling of the 
supercell dimensions and atomic positions according to the new
theoretical equilibrium lattice parameter $a_0^{SCF}$=5.47 \AA\ 
of c-Si. We also perform 
an additional optimization of the internal degrees of freedom 
through minimization of the total-energy and atomic forces, 
and finally we use the resulting configuration
for the  calculation of the  electronic--structure.
Because of the non-periodicity in the amorphous 
slab (see Fig. \ref{f:potential}), 
it is not possible to define an average electrostatic potential 
in that region and therefore  to extract the band alignment according to 
the usual procedure explained in Ref. \cite{review_intf} for the
crystalline junctions. Instead, from the
supercell calculations we can extract the Fermi energy $E_F$ and the 
average 
electrostatic potential in the crystalline region, $\langle V\rangle_c$,
both measured
with respect to the average electrostatic potential of the entire supercell
which is set to zero
(see Fig. \ref{f:sb}). From the electronic structure
calculations of bulk c-Si we obtain the position of its valence band top
edge $E_{v,c}$ 
with respect to its own average electrostatic potential, and therefore 
we can  calculate the band alignment
in terms of the Schottky barrier: $\phi_p=E_F-E_{v,c}-\langle V\rangle_c$.
The {\em numerical} uncertainty associated
to $\phi_p$, coming mainly from the determination of  $E_F$, is
of the order of 0.1 eV. Calculating  the band alignment in terms of 
$\phi_p$ is consistent also with the fact that
we treat the a-Si/c-Si interface as a metallic system,
due to the presence---in general---of defect-induced gap states.

We discuss the results for 6 different a-Si/c-Si configurations,
obtained by means of different details in the above preparation procedure.
Variations on such a protocol were aimed at producing a-Si samples 
with varying concentration of coordination defects.
In  Tab. I we report the relative  occurrence in the amorphous part
of under-coordinated and over-coordinated Si atoms,
calculated simply using a bond-cutoff distance 15\% larger than the 
crystalline bond length. This choice is consistent with the actual
position of the first minimum of the pair correlation function computed in
the amorphous slab.
The six samples are representative of at least three quite different 
atomic--scale arrangements, with  very low (sample A), intermediate
(B, C) and high (D, E, F) concentration of defects. Such a diversity of
relative occurrence of coordination defects also reflects in the 
absolute average atomic coordination
that varies from  3.99 for sample A (small under--coordination) to
4.3 for sample F (sizeable over--coordination). These values
are consistent with experimental findings on real a-Si samples~\cite{servalli}.
Within the last two categories (samples B, C and D, E, F, respectively), 
we present the results for two or three samples in order to 
examine the possible influence of fine structural details.
The cell dimension $c/a$ is slightly varying in the different samples,
with positive and negative small deviations from the ideal value $c/a$=5. 

{\it Results for defect-free a-Si/c-Si samples.} --
We first focus on sample $A$, which, at variance with our typical 
procedure,
has been obtained starting directly from a 320-atom supercell
and not joining two separate a-Si and c-Si cells as described in
step $(i)$.
The amorphization of the a-Si slab has been achieved by the 
WWW model \cite{WWW}; then the sample has been treated with
EDIP--MD for the volume optimization, and with 
a TB {\em athermic} annealing for a final optimization of the atomic 
positions. Sample $A$
has only two three-fold coordinated defects, one at each interface.
It has a clear semiconducting character, with a well defined energy gap,
and a band alignment   $\phi_p$=0.06 eV.
In this case it is possible to calculate also the VBO,
since the position of the topmost valence band edge in the amorphous slab 
with respect to $E_F$ is clearly obtained from supercell calculations. 
We found VBO=0.04 eV, with 
c-Si higher, sizeably smaller than the predictions of
Refs.~\cite{VdW,Allan} (0.25 eV and 0.19 eV respectively). The discrepancy
may be ascribed to the fact that those calculations do not consider 
directly
the interface; in Ref.~\cite{Allan}, for instance, the 
electrostatic potential lineup is {\it assumed} to be zero.

Because of its constituents, sample $A$ can be regarded as an isovalent 
lattice-matched semiconductor/semiconductor crystalline heterojunction.
Indeed it obeys the general trend of this class of junctions, where the VBO 
is a bulk-related effect, independent on the interface details 
\cite{review_intf}: remarkably,  the two interfaces in the supercell,
although slightly morphologically different, are really {\it equivalent} 
in terms of electronic structure, since {\it no electric fields} are 
present 
in the bulk slabs.

This behavior is consistent  with  experimental findings for the analogous 
system a-Si:H/c-Si \cite{Evangelisti},
where a constant VBO=0.44 eV has been found  for
different samples characterized by the same amount of H but
different thickness of the interface region.\cite{note}

{\it Band alignment at Si/Si crystalline junctions.} --
In order to examine the role played by the density and the 
crystalline symmetry on the band alignment,  we consider two particular 
crystalline Si junctions: the unstrained diamond/strained diamond
 and the heterocrystalline cubic diamond/hexagonal diamond interfaces.
For the former case we consider specifically
a positive strain of 1\% along the (111) direction,
for a direct comparison with the case of hexagonal diamond
which {\it naturally} exhibits such strain: we found
VBO=0.05$\pm$0.01 eV (with the strained slab
higher) after optimization of the internal degrees of freedom in the 
strained Si  slab.
The contribution to the VBO 
due to the change in density is 0.02 eV,  related to the
{\it hydrostatic} deformation potential \cite{def-pot}. 
The remaining contribution 
comes from the strain-induced {\em splitting} of the valence band top edge 
(0.10 eV).

We simulate the heterocrystalline  cubic diamond/hexagonal diamond Si
junction with a supercell with 30 atoms, namely with 3 unit slabs
for both phases along the (111) direction \cite{nota}.
After a  careful optimization of the macroscopic 
and microscopic degrees of freedom, which  gives for the structure of the
hexagonal diamond results close to those of Ref.\cite{wz-zunger}, we found
VBO=0.23$\pm$0.01 eV, 
with hexagonal diamond higher than  cubic diamond, as in Ref.\cite{wz-jap}.
Remarkably, the VBO is much larger than in the previous junction,
although the change in the density between the constituent slabs is the
same.
The largest contribution to the VBO comes from the fact that
in the  hexagonal diamond the valence band width is larger   than in cubic
diamond, and the valence top has a  splitting of 0.30 eV.

The common feature of the semiconducting a-Si/c-Si interface and of 
crystalline
Si/Si junctions is the reduced symmetry of one of the constituent slabs:
this can be indeed considered as the origin of the 
VBO in these systems, through a splitting of 
the valence band top edge manifold, more pronounced for the structure
having the largest anisotropy (hexagonal diamond).

{\it Results for other a-Si/c-Si samples} --
We discuss now the results obtained for other a-Si/c-Si samples, which are
characterized by a concentration of defects in the a-Si region at least 
of the order of  $\sim$10\% and 
by a semimetallic behavior due to defect-induced gap states.
Sample $B$ is obtained following the same procedure for sample $A$, except
that the starting configuration is obtained joining two separate cells
(c-Si and a-Si, 
treated with the WWW model \cite{WWW});
sample $C$ is obtained without the use of the WWW model, and
following exactly all the steps $(i)-(iv)$ previously listed,
with a slow quench rate in step $(i)$ in order to keep 
the number of defects very low; 
sample $D$ is obtained with a higher quenching rate
in step $(i)$ and a TB {\em finite-temperature} annealing in step 
$(iv)$;
sample $E$ with a TB {\em athermic} annealing; finally, sample $F$
with with the use of EDIP--MD only
(i.e., steps $(iii)-(iv)$ were skipped).

Our key-result is that for all the samples from $B$ to $F$
the band alignment  has almost the same value $\phi_p$=0.22 eV,
independent on structural  details  such as concentration and 
distribution of defects and  density of the amorphous slab.
The average 
electrostatic potential in the crystalline slab $\langle V\rangle_c$
is very sensitive to the defect concentration, and infact the values
for the samples with intermediate defect concentrations (B, C) are very
different from those with high defect concentrations (D, E, F), even 
by 0.48 eV (see Tab. I). 
Nonetheless, the Schottky barrier $\phi_p$ varies within a very small range 
(of 0.04 eV) for {\em all} these samples.

This result for  defect-rich samples
is consistent with the general trend
reported for the semiconductor/metal junctions \cite{review_intf}, i.e. 
that the Schottky barrier is poorly sensitive to the metal type
but depends on the semiconductor.  
In fact, what changes from sample $B$ to $F$ is only 
the highly defected a-Si slab,  which  acts as the ``metallic'' 
constituent. 
Moreover, since c-Si is homopolar, neither
interface-specific details could change the band alignment,
as instead it may occurs in case of heterovalent semiconductor 
heterostructures and metal/polar-semiconductor junction. The independence
on interface details is also numerically verified from our calculations:
in all our samples  electric fields
within the c-Si slabs are absent, whereas they 
would be present if the two interfaces within the supercell
were non-equivalent.

{\it Conclusions.} - 
Present results for amorphous/crystalline Si interfaces can be rationalized 
along the findings of the linear-response theory of the
band alignment at semiconductor interfaces \cite{review_intf}
and summarized as follows: $(i)$
the  band alignment is almost vanishing for defect-free samples 
and sizeable in the case of defect-rich samples; 
$(ii)$ in the latter case it is independent on the particular defect
concentration and distribution: such result can be
 ascribed to the semi-metallic behavior
of the amorphous region, and explained by analogy with the case of
Schottky barriers which are almost independent on the metal type; 
$(iii)$ in any case
it is not an  interface-specific effect, but  rather it is 
a bulk-related effect; $(iv)$ 
by inspection of the different 
a-Si/c-Si samples and for comparison with other  Si/Si crystalline 
junctions,
the largest contribution to the band alignment
is not due to changes in the density but rather to
 a broadening of the valence band
and a  splitting of its top manifold which typically occurs in the phases
of reduced symmetry.

This work was supported by the INFM Parallel Computing Initiative.


\narrowtext
\begin{table}[h]
\caption{Main structural and electronic properties of the six different 
a-Si/c-Si samples considered in this work. The percentages of atoms 
in the amorphous part which are under-coordinated ($T_{NN<4}$)
and over-coordinated ($T_{NN>4}$), where NN means ``nearest
neighbors'', are indicated in the second and third columns.}
\begin{tabular}{lrrccrc}
Sample& $T_{NN<4}$& $T_{NN>4}$
&$c/a$& $E_F$ (eV)&$\langle V\rangle_c$ (eV)&$\phi_p$ 
(eV)\\ 
\hline
A & 1\%~ & 0\%~ &  5.039 & 6.26 & $-$0.11~~ & 0.06 \\
B & 7\%~ & 6\%~ &  5.126 & 6.22 & $-$0.30~~ & 0.20 \\ 
C & 0.5\%~ & 9\%~ &  5.100 & 6.33 & $-$0.24~~ & 0.24 \\ 
D & 2\%~ & 36\%~ &  4.922 & 6.71 & 0.18~~ & 0.21 \\ 
E & 3\%~ & 27\%~ &  4.922 & 6.70 & 0.15~~ & 0.23 \\ 
F & 0.5\%~ & 36\%~ &  4.922 & 6.68 & 0.14~~ & 0.22 
\end{tabular}
\label{tab.alat}
\end{table}

\narrowtext

\begin{thebibliography}{99}
\bibitem{nota}
The unit slab of 
cubic diamond in the (111) direction has 6 atoms, the one of hexagonal
diamond has 4 atoms, and coincides with its minimum unit cell.
\end{thebibliography}
\end{document}